\documentclass[sigconf]{acmart}
\AtBeginDocument{%
  \providecommand\BibTeX{{%
    \normalfont B\kern-0.5em{\scshape i\kern-0.25em b}\kern-0.8em\TeX}}}
   
\usepackage{setspace}
\usepackage{subcaption}
\usepackage[justification=centering]{caption}
\usepackage{booktabs} 
\usepackage{float}
\usepackage{graphicx}
\usepackage{amsmath,amsthm} 
\usepackage{hyperref}
\usepackage{array}
\usepackage{tikz}
\usepackage{tikzscale}
\usepackage{xspace}
\usepackage{pgfplots}
\usepackage{enumitem}
\usepgfplotslibrary{groupplots}
\pgfplotsset{compat=1.16}
\usepackage[normalem]{ulem}
\usepackage{soul}

\newcommand{\Name}{\textsf{{Bingo}}\xspace}

\acmSubmissionID{mobiwac13s} 
\copyrightyear{2021}
\acmYear{2021}
\setcopyright{acmcopyright}\acmConference[MobiWac '21]{Proceedings of the 19th ACM International Symposium on Mobility Management}{November 22--26, 2021}{Alicante, Spain}
\acmBooktitle{Proceedings of the 19th ACM International Symposium on Mobility Management (MobiWac '21), November 22--26, 2021, Alicante, Spain}
\acmPrice{15.00}
\acmDOI{10.1145/3479241.3486694}
\acmISBN{978-1-4503-9079-8/21/11}

\begin{document}

\fancyhead{}
\settopmatter{authorsperrow=4}

\title{Social Groups Based Content Caching in Wireless Networks}

\author{Nimrah Mustafa}
\affiliation{%
  \institution{LUMS University}}
\email{18030049@lums.edu.pk}

\author{Imdad Ullah Khan}
\affiliation{%
  \institution{LUMS University}}
\email{imdad.khan@lums.edu.pk}

\author{\hspace{-.05in}Muhammad Asad Khan}
\affiliation{%
  \institution{Hazara University}}
\email{asadkhan@hu.edu.pk}

\author{Zartash Afzal Uzmi}
\affiliation{%
  \institution{LUMS University}}
\email{zartash@lums.edu.pk}

\begin{abstract}

The unprecedented growth of wireless mobile traffic, mainly due to multimedia traffic over online social platforms has strained the resources in the mobile backhaul network. A promising approach to reduce the backhaul load is to proactively cache content at the network edge, taking into account the overlaid social network. Known caching schemes require complete knowledge of the social graph and mainly focus on one-to-one interactions forgoing the prevalent mode of content sharing among circles of `friends'. We propose \Name, a proactive content caching scheme that leverages the presence of interest groups in online social networks. The mobile network operator (MNO) can choose to incrementally deploy \Name at select network nodes (base stations, packet core, data center) based on user profiles and revenue numbers. We approximate the group memberships of users using the available user-content request logs without any prior knowledge of the overlaid social graph. \Name can cater to the evolving nature of online social groups and file popularity distribution for making caching decisions. We use synthetically generated group structures and simulate user requests at the base station for empirical evaluation against traditional and recent caching schemes. \Name achieves up to $30\%-34\%$ gain over the best baseline.

\end{abstract}



\maketitle

\section{Introduction}\label{introduction}

The rapid growth in the use of smart mobile devices over the last decade has resulted in an explosive increase in mobile data traffic. The major contributing factor is the increasing number of one-to-many transmissions in the form of multimedia messaging to groups and content posted with fans over social networks \cite{bastug2014living,shanmugam2013femtocaching,yang2018locass}. The resulting redundant and repetitive transmission of content \cite{ Zhang2018Hierarchical} stresses the cellular backhaul.

Proactively caching content at the network edge and \emph{locally} serving repeat requests will alleviate the wireless backhaul resource stress and also reduce user-perceived latency. Traditional web caching systems do not work as well in modern-day traffic where an enormous amount of content is created, shared, and forwarded by a large number of users. Online Social networks are often organized in the form of overlapping \emph{interest groups} followers of Facebook pages, Twitter users, and members of Instagram circles. These groups or communities are characterized by their similar interests (usually requesting the same contents). 

This paper proposes \Name, a social community-based edge caching scheme. Community structure in the (logical) social network is not known to the MNO, \Name estimates the groups from the user-request log. \Name leverages this {\em social locality} information (contents shared with social groups) to proactively cache content at the network edge (e.g. base station). Thus, multiple members of a single social group receive the content of their mutual interest from the respective edge nodes. \Name outperforms traditional caching schemes in terms of the cache hit ratios, thus resulting in (i) reduced usage of backhaul bandwidth, and (ii) smaller latency for end-users.

\begin{figure}[htb!]
        \includegraphics[page=3,width=.8\linewidth]{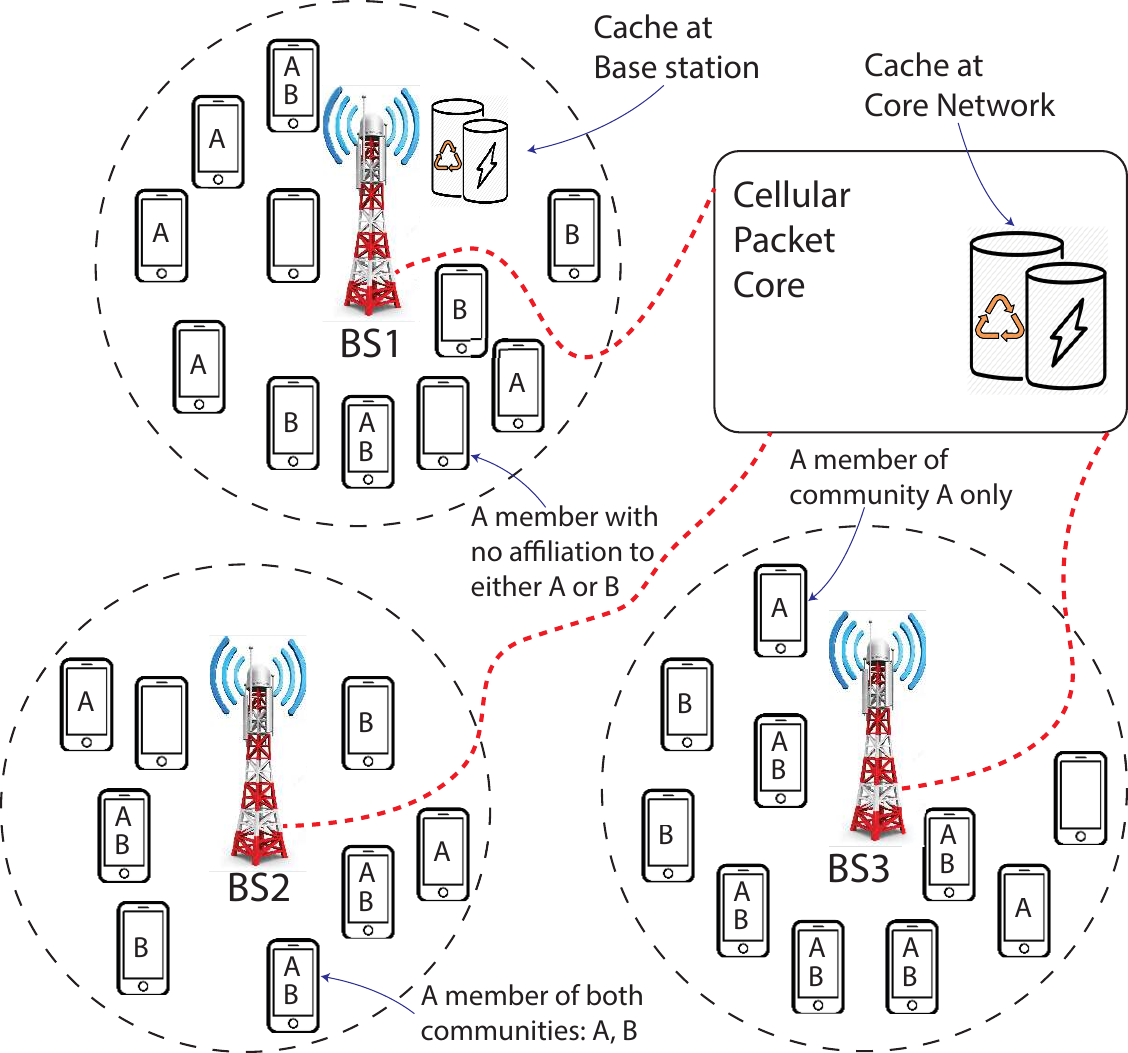}
		\caption{Users belonging to two overlapping communities $A$ and $B$ are scattered in three cells served by base stations $BS1$, $BS2$, and $BS3$. Possible caching locations are at the base stations (e.g., at $BS1$, as shown) and at the cellular core.}
		\label{overlap-multiple-bs}
\end{figure} 

A practical feature of \Name is its flexibility to be deployed incrementally at multiple levels of regional granularity, as per the business priorities of the mobile operators. An operator may choose to deploy \Name at a single base station (e.g., a cell containing $BS1$ in Figure~\ref{overlap-multiple-bs}). This will allow content caching at the base station. For instance, if the content shared with community $A$ is cached at the $BS1$, it will be served locally from this edge cache to the six members of community $A$ served by $BS1$. 

Alternately, \Name may be deployed in the cellular packet core of a provider, allowing content caching in the core, bringing caching benefit to the members of a social community split into multiple base stations. In Figure~\ref{overlap-multiple-bs}, if the content destined for community $A$ is cached at the cellular packet core, then all 18 members of community $A$ across the three cells will benefit from this \emph{locality of reference}. \Name also supports a cache hierarchy by simultaneous deployment at select base stations, the packet core, and a remote data center. With hierarchical deployment, the decision engine in \Name will cache a piece of content only at those edge nodes which may serve multiple members of the social community interested in that content.
Many known caching schemes take into account the overlaid logical social network to decide \emph{What to cache} and \emph{Where to cache}.  There also exist collaborative caching schemes using device-to-device (D2D) communication that cache content at users' devices not just for themselves but also for their ``social friends'' in a \emph{tit for tat} manner. These schemes however require complete knowledge of the social graph, focus on one-to-one interactions rather than content sharing among circles of {\em friends}, and/or do not adapt to quick content popularity changes. 

\Name addresses all the above concerns: it does not require knowledge of the social graph; it does not even try to estimate the social graph (respecting the security and privacy of user traffic). Instead, employing existing overlapping community detection techniques on the weighted user network modeled from the request log, \Name approximates the community structure in the overlaid social network. For making a caching decision, \Name considers local (instead of global) popularity of content---by measuring the interest of relevant communities. Finally, \Name also incorporates other features of social traffic such as {\em geographic locality} (content shared among users that are geographically close by) and {\em temporal locality} (recent content is more popular) in caching decisions.

We empirically evaluate \Name in an extensive set of experiments with varying community structures, a wide range of network densities, and file popularity distributions. All parameters used for generating synthetic test data are set to values observed in real-life data and as used in the literature. Our empirical evaluation demonstrates that \Name achieves up to $34\%$ gain over known edge caching schemes in terms of cache hit-ratio.

Altogether, this paper makes the following contributions:
\begin{itemize}
	\item Methodology to estimate the (evolving) social communities based on past user requests without additional information
	\item An open-source prototype implementation of \Name (including community detection, community identification, and caching decision engine) \footnote{\url{https://github.com/NimrahMustafa/SocialCommunityBasedCaching.git}}
	\item A thorough evaluation of \Name  using synthetic social groups on a broad range of parameters 
\end{itemize}

The rest of the paper is organized as follows. We review existing literature in Section \ref{related_work} and explain the algorithm in Section \ref{description}. We present the experimental setup in Section \ref{experimental_setup} and discuss empirical results in Section \ref{Results}, concluding the paper in Section \ref{conclusion}.

\section{Related Work}\label{related_work}

Content caching has evolved as an integral part of wireless networks. Edge caching in wireless networks employs data and social network analysis and machine learning methods for estimating content popularity~\cite{bernardini2013mpc}, determining request frequency~\cite{ong2014fgpc}, optimal content placement~\cite{bastug2014living}, and collaborative caching~\cite{yang2018locass} between end-users by storing content at user devices for later serving on-demand using the D2D communications~\cite{Ji2016WirelessD2D, Golrezaei2014BSAssistedD2D}.

The exact location of the cache plays an important role in the effectiveness of a caching scheme. Content can be cached at the base stations~\cite{Poularakis2014Multicast, Poularakis2014Approximation}, special purpose femtocells~\cite{shanmugam2013femtocaching}, or in a hierarchical storage system distributed across the mobile network~\cite{Tran2017Cooperative, Zhang2018Hierarchical}. 

There has been recent research interest in leveraging personal and contextual information to determine what content to cache~\cite{Muller2017Context}. Such human-centric information includes visited locations, gender, job, and device type and other user attributes ~\cite{ali2021predicting}. User mobility patterns are used in~\cite{Zhang2019MobilityAware}) to pick the best location to serve content from a distributed cache.

Social network information has been utilized to enhance caching performance. In~\cite{bastug2014living} influential nodes are identified as caching locations that will serve their social friends via D2D communications. A more recent approach for D2D caching exploits the knowledge of pairwise social interactions~\cite{Weifeng2020CachingSocialTrust}. In~\cite{yang2018locass, Zhu2017SociallyMotivated} a game-theoretic approach is proposed for mobile users to cache contents, both for themselves and for their friends.

\section{Description}\label{description}
We describe \Name at the level of a single cell. Owing to the hierarchical and incremental deployability of \Name, the single-cell model readily extends to caching in the core or a remote datacenter.

\noindent \textbf{Network Model:} Consider a base station, connected to the core network with a backhaul link, serving $N$ users $U=\{u_1, u_2, ....,u_N\}$ which are organized into $K$ possibly overlapping communities $\mathcal{C}=\{C_1, C_2,\ldots,C_K\}$, $C_i \subseteq U$. Users request files from a library  $\mathcal{F} =\{f_1, f_2,...,f_F\}$, where each file $f_j$ has an associated popularity $p_j \in (0,1)$. The community structure $\mathcal{C}$ and $p_j$ are not known at the base station. Cache storage with the capacity to store $S$ files is installed at the base station. \textsc{wlog}, we assume that all files have the same size and a file is either stored completely or not at all \cite{shanmugam2013femtocaching}.

\noindent  \textbf{Key Idea:} The guiding principle for caching in \Name is that if a file is accessed by a member of some community, then almost the entire community would (ultimately) request it. In this regard, users with similar access patterns (requesting the same content within the same time frame) are considered to be a community. This notion of a community is only for the internal working of the estimation algorithm and does not have any implications on the formation of communities, where membership of users is a direct result of social ties, common interests, and event participation.
  
 \noindent \textbf{Process Overview:} The continuous stream of requests is divided into chunks each containing $L$ consecutive requests. All requests arriving in a chunk $D_k$ are logged and used to model a user network to estimate the community structure $\mathcal{C}'_{k+1}$ for the next chunk $D_{k+1}$. This model incorporates the formation of new and dissolution of old communities, as well as the members joining or leaving existing communities. The chunk size $L$ is defined as the number of requests, instead of time elapsed, to cater for traffic variability and ensure that sufficient requests are available to reliably estimate the community structure and can be altered to accommodate the frequency of significant changes in the community structure. With the estimated community structure $\mathcal{C}'_k$ in place, for each request $(u_i,f_j)$ in $D_k$, since $u_i$ may belong to many communities, we identify the community as a member of which, $u_i$ has requested $f_j$ after which a caching decision for $f_j$ is made and an existing file in the cache may be removed as per the eviction policy. If cached, remaining requests by users of the community can be served locally. 
  
\noindent \textbf{Community Detection: } To estimate the community structure, we first transform the request log into an undirected weighted graph to model a user-user network. The users form the nodes, and an edge between any pair of users is weighted by the number of common files requested by both users. Edges with weight below $\beta$ are dropped. The threshold $\beta$ can be appropriately set to reduce the likelihood of incorrect inclusion or exclusion of a user in a community, i.e. for a given period, its value should be high enough for a low likelihood of two users belonging to different communities requesting at least $\beta$ and low enough for a high likelihood of members of a community requesting at least $\beta$ files. Then, an existing community detection algorithm can be directly run on the graph. We use a simple conductance-based algorithm \cite{communitydetection} for community detection. Although community detection is computationally expensive, it is sufficient to periodically perform this task in the background during off-peak hours, since these communities evolve rather slowly. \Name would cater to significant changes occurring in a period on the order of hours.

\begin{figure}[tb!]
        \centering
		\includegraphics[page=1,scale=0.9]{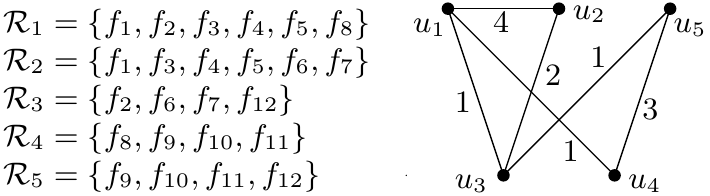}
		\caption{Modelling request log as a weighted graph. $\mathcal{R}_i$ is the set of files requested by $u_i$}
		\label{transaction_ex}
\end{figure} 

\noindent \textbf{Community Identification: } A minimum of $\xi$ requests by users for a file $f_j$ are accumulated before a community is identified, which must be among the set of common communities to which users who have all requested $f_j$ belong. The optimal value of $\xi$ can be fine-tuned depending on the overlap factor of the communities which, in many real-world social networks, is small, communities can be identified accurately in a short time, i.e. for a small value of $\xi$. We must account for users who may have individually requested $f_j$ and may not have any common communities with the other requesting users since, otherwise, the likelihood of the intersection of communities to which requesting users belong being empty will be higher. To this end, we sort the requesting users in descending order by the number of communities they belong to. The intersection process begins with the user belonging to the most number of communities. In each step, the intersection of the set of potentially identifiable communities is taken with the set of communities to which the next user belongs. This process stops when some minimum number of potential communities $\rho$, a configurable parameter, remain from which the largest is selected as the identified community $\chi_j$. 

\begin{figure}[tb!]
	\begin{center}
		\includegraphics[page=2,scale=0.9]{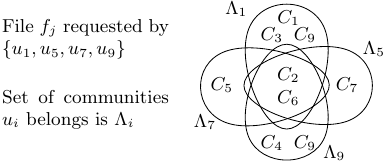}
		\caption{Identified community $\chi_j$ is the larger of $C_2$ and $C_6$}
		\label{commiden_ex}
	\end{center}
\end{figure} 

\noindent \textbf{Caching Decision: }The caching decision for a file $f_j$ is primarily driven by $|\chi_j|$, i.e. the number of users belonging to the identified community. We assign each requested file $f_j$ a score $s_j$ and cache the highest-scoring files. If a community is identified, $s_j=|\chi_j|-\xi$. Otherwise, $s_j=1$. For each user who belongs to $\chi_j$ and requests $f_j$ after $\chi_j$ is identified, $s_j$ is decremented. For each cached file $f_j$, the number of requests elapsed since $f_j$ was last requested $l_j$ is used to incorporate a recency factor in the caching decision to avoid unnecessarily occupation of cache space. If $f_j$ is evicted without having served all its expected requests, the number of served requests $r_j$ is temporarily retained so when $f_j$ is next cached with identified community $\chi_j$, $s_j$ is set based on $|\chi_j|-r_j$ instead of $|\chi_j|$ which is a higher score than $f_j$ deserves and would adversely impact caching of more deserving files. The record is deleted once the expected number of requests of $f_j$ have been served. The cache is modeled as a Min-Heap to support the eviction policy, i.e., find and delete the file with the minimum score.

\begin{figure*}[!hbt]
\centering
\includegraphics{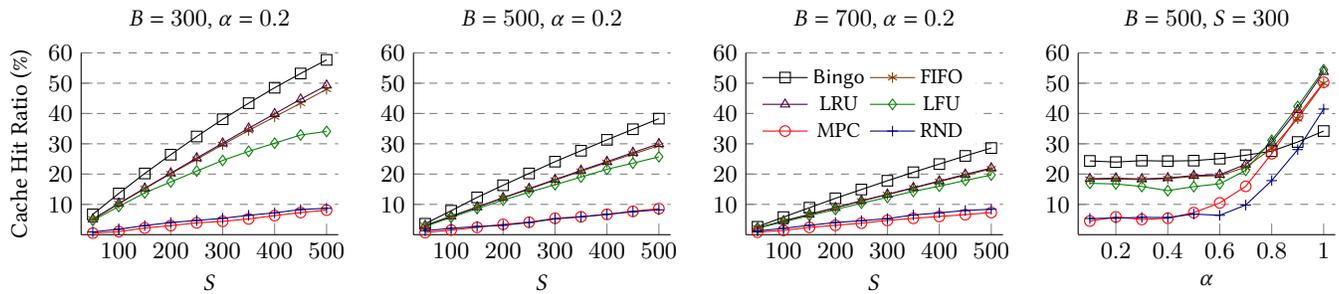}
\caption{Varying Cache Capacity $S$, Batch Size $B$ (controls traffic volume) and ZipF parameter $\alpha$ (controls file popularity)}
\label{varyCacheCapWithBatchSize}
\end{figure*}

\section{Experimental Setup}\label{experimental_setup}

MSNs have two components: the social network among users and the mobile network of user requests for content. Since the ownership of these two components lies with different entities, the combination of both social and mobile data traces is limited. Due to this lack of real data, we simulate request arrivals at the base station using both synthetic and real-world community structures. 

Since it is well-known most social networks (Twitter, Facebook, etc.) are scale-free, we use the affiliation graph model (AGM) \cite{Yang2012Community} to construct the social community structure. We chose the AGM model over other benchmark community generation models such as LFR, which generate {\it user-user} networks since a {\it user-community} bipartite membership network suffices for our purpose as we do not use the pairwise links between users. We detail the process of realistic request arrival simulation in the source code.

Users request files according to their popularity, which is sampled from the Zipf distribution, which is well known for modeling file popularity. \cite{bastug2014living}. Popularity $p_j$ of a file $f_j$ is defined as $p_j = \frac{j^{-\alpha}}{\sum_{l \in {\mathcal F}}\; l^{-\alpha}}$, where $\alpha \geq 0$ is the steepness parameter of the popularity curve. A smaller value of $\alpha$ implies that fewer files are more popular. ${\mathcal F}$ is a static set of $10^6$ files (effectively infinite). Parameters for community estimation and identification are set as $\beta=3$, $\xi=4$, and $\rho=2$ empirically with no notable deviation from expectation with setting to other values. We evaluate the caching engine using the known community structure to avoid any bias of the community detection algorithm. Since community detection is a widely studied problem in itself, more sophisticated algorithms can be employed to improve performance.

We compare the cache hit ratio, which directly translates to the traffic volume offloaded from the backhaul links, with five baseline schemes - FIFO, LFU, LRU, MPC, and Random (RND) Caching. Since our focus is on \emph{What to cache}, no meaningful comparison can be made with other state-of-the-art works which primarily focus on \emph{Where to cache}. Other quality metrics such as page load times and backhaul bandwidth usage are easily derived from hit ratio when network specifications (such as link capacities) are available. The impact on the hit ratio of three key system parameters - traffic volume, cache capacity, and content popularity, is studied. 

\section{Results and Discussion }\label{Results}

Since a series of batches of requests is used to simulate their arrival, the traffic volume is controlled by the parameter batch size $B$ that indicates the number of currently active communities (for which users are requesting content). As expected, the hit ratio decreases for all caching schemes with an increase in traffic volume. However, \Name maintains a significant gain over the baselines, up to $34\%$ over the closest baseline, i.e. LRU. When traffic volume in the network is higher, a better cache hit ratio becomes more crucial to reduce response latency and backhaul load. We also observe that increasing the cache capacity $S$ increases the performances gap and thus it is more fruitful for \Name compared to the baseline schemes.

Content popularity is controlled by the parameter $\alpha$ of the Zipf distribution. The hit ratio increases slightly for all caching schemes when $\alpha$ is increased, i.e. file popularity becomes less uniform because as the same popular content is being requested more often (by more communities), more requests are served locally as compared to when fewer files were popular, i.e. more diverse content was being requested. Our experiments show that for $\alpha>0.8$, MPC, LFU, and LRU show a drastic increase in hit ratio whereas \Name shows a relatively smaller increase. However, such skewed distributions are not realistic and do not reflect request patterns in the real world. \Name consistently outperforms the baselines for practical popularity distributions~\cite{yang2018locass,bastug2014living}. 

Note that \Name outperforms MPC, the main competitor since the number of files available is essentially infinite, and \textit{caching what more users are expected to request is better than expecting users to request only a few popular files}. Furthermore, this implementation of MPC uses exactly known file popularity distribution, unlike in reality. We consider static popularity distributions since \Name does not depend on the global popularity of the content and instead draws on the local dynamics of `popularity'.  

\section{Conclusion}\label{conclusion}
We proposed \Name, a proactive edge caching scheme for cellular networks that utilizes structural information of the social network. For each requested content, we identify the approximate community that maintains the maximal interest in that content. The estimated size of the community, together with the current cache status is used to make caching decisions.
MNO may choose to deploy \Name at select sites based on expected revenues. We generate user requests using synthetic communities, which simulate real-world scenarios, to empirically evaluate \Name. We demonstrate that \Name substantially outperforms the (more) reactive caching schemes on varying network traffic volumes and community structures.

\bibliographystyle{ACM-Reference-Format}
\bibliography{Bibliography_Content_Caching}


\begin{thebibliography}{18}


\ifx \showCODEN    \undefined \def \showCODEN     #1{\unskip}     \fi
\ifx \showDOI      \undefined \def \showDOI       #1{#1}\fi
\ifx \showISBNx    \undefined \def \showISBNx     #1{\unskip}     \fi
\ifx \showISBNxiii \undefined \def \showISBNxiii  #1{\unskip}     \fi
\ifx \showISSN     \undefined \def \showISSN      #1{\unskip}     \fi
\ifx \showLCCN     \undefined \def \showLCCN      #1{\unskip}     \fi
\ifx \shownote     \undefined \def \shownote      #1{#1}          \fi
\ifx \showarticletitle \undefined \def \showarticletitle #1{#1}   \fi
\ifx \showURL      \undefined \def \showURL       {\relax}        \fi
\providecommand\bibfield[2]{#2}
\providecommand\bibinfo[2]{#2}
\providecommand\natexlab[1]{#1}
\providecommand\showeprint[2][]{arXiv:#2}

\bibitem[\protect\citeauthoryear{Ali, Shakeel, Khan, Faizullah, and Khan}{Ali
  et~al\mbox{.}}{2021}]%
        {ali2021predicting}
\bibfield{author}{\bibinfo{person}{S. Ali}, \bibinfo{person}{M.~H. Shakeel},
  \bibinfo{person}{I. Khan}, \bibinfo{person}{S. Faizullah}, {and}
  \bibinfo{person}{M.~A. Khan}.} \bibinfo{year}{2021}\natexlab{}.
\newblock \showarticletitle{Predicting attributes of nodes using network
  structure}.
\newblock \bibinfo{journal}{\emph{ACM Transactions on Intelligent Systems and
  Technology}} \bibinfo{volume}{12}, \bibinfo{number}{2}
  (\bibinfo{year}{2021}), \bibinfo{pages}{1--23}.
\newblock


\bibitem[\protect\citeauthoryear{{Bastug}, {Bennis}, and {Debbah}}{{Bastug}
  et~al\mbox{.}}{2014}]%
        {bastug2014living}
\bibfield{author}{\bibinfo{person}{E. {Bastug}}, \bibinfo{person}{M. {Bennis}},
  {and} \bibinfo{person}{M. {Debbah}}.} \bibinfo{year}{2014}\natexlab{}.
\newblock \showarticletitle{Living on the edge: The role of proactive caching
  in $5${G} wireless networks}.
\newblock \bibinfo{journal}{\emph{IEEE Communications Magazine}}
  \bibinfo{volume}{52}, \bibinfo{number}{8}, \bibinfo{pages}{82--89}.
\newblock


\bibitem[\protect\citeauthoryear{{Bernardini}, {Silverston}, and
  {Festor}}{{Bernardini} et~al\mbox{.}}{2013}]%
        {bernardini2013mpc}
\bibfield{author}{\bibinfo{person}{C. {Bernardini}}, \bibinfo{person}{T.
  {Silverston}}, {and} \bibinfo{person}{O. {Festor}}.}
  \bibinfo{year}{2013}\natexlab{}.
\newblock \showarticletitle{\uppercase{MPC}: Popularity-based Caching Strategy
  for Content Centric Networks}. In \bibinfo{booktitle}{\emph{IEEE
  International Conference on Communications}}. \bibinfo{pages}{3619--3623}.
\newblock


\bibitem[\protect\citeauthoryear{{Golrezaei}, {Mansourifard}, {Molisch}, and
  {Dimakis}}{{Golrezaei} et~al\mbox{.}}{2014}]%
        {Golrezaei2014BSAssistedD2D}
\bibfield{author}{\bibinfo{person}{N. {Golrezaei}}, \bibinfo{person}{P.
  {Mansourifard}}, \bibinfo{person}{A. {Molisch}}, {and} \bibinfo{person}{A.
  {Dimakis}}.} \bibinfo{year}{2014}\natexlab{}.
\newblock \showarticletitle{Base-Station Assisted Device-to-Device
  Communications for High-Throughput Wireless Video Networks}.
\newblock \bibinfo{journal}{\emph{IEEE Transactions on Wireless
  Communications}} \bibinfo{volume}{13}, \bibinfo{number}{7},
  \bibinfo{pages}{3665--3676}.
\newblock


\bibitem[\protect\citeauthoryear{{Ji}, {Caire}, and {Molisch}}{{Ji}
  et~al\mbox{.}}{2016}]%
        {Ji2016WirelessD2D}
\bibfield{author}{\bibinfo{person}{M. {Ji}}, \bibinfo{person}{G. {Caire}},
  {and} \bibinfo{person}{A. {Molisch}}.} \bibinfo{year}{2016}\natexlab{}.
\newblock \showarticletitle{Wireless Device-to-Device Caching Networks: Basic
  Principles and System Performance}.
\newblock \bibinfo{journal}{\emph{IEEE Journal on Selected Areas in
  Communications}}  \bibinfo{volume}{34}, \bibinfo{pages}{176--189}.
\newblock


\bibitem[\protect\citeauthoryear{{Lu}, {Sun}, {Wen}, {Cao}, and {Porta}}{{Lu}
  et~al\mbox{.}}{2015}]%
        {communitydetection}
\bibfield{author}{\bibinfo{person}{Z. {Lu}}, \bibinfo{person}{X. {Sun}},
  \bibinfo{person}{Y. {Wen}}, \bibinfo{person}{G. {Cao}}, {and}
  \bibinfo{person}{T. {Porta}}.} \bibinfo{year}{2015}\natexlab{}.
\newblock \showarticletitle{Algorithms and Applications for Community Detection
  in Weighted Networks}.
\newblock \bibinfo{journal}{\emph{IEEE Transactions on Parallel and Distributed
  Systems}} \bibinfo{volume}{26}, \bibinfo{number}{11},
  \bibinfo{pages}{2916--2926}.
\newblock


\bibitem[\protect\citeauthoryear{{Müller}, {Atan}, {van der Schaar}, and
  {Klein}}{{Müller} et~al\mbox{.}}{2017}]%
        {Muller2017Context}
\bibfield{author}{\bibinfo{person}{S. {Müller}}, \bibinfo{person}{O. {Atan}},
  \bibinfo{person}{M. {van der Schaar}}, {and} \bibinfo{person}{A. {Klein}}.}
  \bibinfo{year}{2017}\natexlab{}.
\newblock \showarticletitle{Context-Aware Proactive Content Caching With
  Service Differentiation in Wireless Networks}.
\newblock \bibinfo{journal}{\emph{IEEE Transactions on Wireless
  Communications}} \bibinfo{volume}{16}, \bibinfo{number}{2},
  \bibinfo{pages}{1024--1036}.
\newblock


\bibitem[\protect\citeauthoryear{{Ong}, {Chen}, {Taleb}, and {Wang}}{{Ong}
  et~al\mbox{.}}{2014}]%
        {ong2014fgpc}
\bibfield{author}{\bibinfo{person}{M. {Ong}}, \bibinfo{person}{M. {Chen}},
  \bibinfo{person}{T. {Taleb}}, {and} \bibinfo{person}{X. {Wang}}.}
  \bibinfo{year}{2014}\natexlab{}.
\newblock \showarticletitle{\uppercase{FGPC}: Fine-grained popularity-based
  caching design for content centric networking}. In
  \bibinfo{booktitle}{\emph{ACM International Conference on Modeling, Analysis
  and Simulation of Wireless and Mobile Systems}}. \bibinfo{pages}{295--302}.
\newblock


\bibitem[\protect\citeauthoryear{{Poularakis}, {Iosifidis}, {Sourlas}, and
  {Tassiulas}}{{Poularakis} et~al\mbox{.}}{2014b}]%
        {Poularakis2014Multicast}
\bibfield{author}{\bibinfo{person}{K. {Poularakis}}, \bibinfo{person}{G.
  {Iosifidis}}, \bibinfo{person}{V. {Sourlas}}, {and} \bibinfo{person}{L.
  {Tassiulas}}.} \bibinfo{year}{2014}\natexlab{b}.
\newblock \showarticletitle{Multicast-aware caching for small cell networks}.
  In \bibinfo{booktitle}{\emph{IEEE Wireless Communications and Networking
  Conference}}. \bibinfo{pages}{2300--2305}.
\newblock


\bibitem[\protect\citeauthoryear{{Poularakis}, {Iosifidis}, and
  {Tassiulas}}{{Poularakis} et~al\mbox{.}}{2014a}]%
        {Poularakis2014Approximation}
\bibfield{author}{\bibinfo{person}{K. {Poularakis}}, \bibinfo{person}{G.
  {Iosifidis}}, {and} \bibinfo{person}{L. {Tassiulas}}.}
  \bibinfo{year}{2014}\natexlab{a}.
\newblock \showarticletitle{Approximation Algorithms for Mobile Data Caching in
  Small Cell Networks}.
\newblock \bibinfo{journal}{\emph{IEEE Transactions on Communications}}
  \bibinfo{volume}{62}, \bibinfo{number}{10}, \bibinfo{pages}{3665--3677}.
\newblock


\bibitem[\protect\citeauthoryear{{Shanmugam}, {Golrezaei}, {Dimakis},
  {Molisch}, and {Caire}}{{Shanmugam} et~al\mbox{.}}{2013}]%
        {shanmugam2013femtocaching}
\bibfield{author}{\bibinfo{person}{K. {Shanmugam}}, \bibinfo{person}{N.
  {Golrezaei}}, \bibinfo{person}{A. {Dimakis}}, \bibinfo{person}{A. {Molisch}},
  {and} \bibinfo{person}{G. {Caire}}.} \bibinfo{year}{2013}\natexlab{}.
\newblock \showarticletitle{Femto\uppercase{C}aching: Wireless content delivery
  through distributed caching helpers}.
\newblock \bibinfo{journal}{\emph{IEEE Transactions on Information Theory}}
  \bibinfo{volume}{59}, \bibinfo{number}{12}, \bibinfo{pages}{8402--8413}.
\newblock


\bibitem[\protect\citeauthoryear{{Tran}, {Hajisami}, and {Pompili}}{{Tran}
  et~al\mbox{.}}{2017}]%
        {Tran2017Cooperative}
\bibfield{author}{\bibinfo{person}{T. {Tran}}, \bibinfo{person}{A. {Hajisami}},
  {and} \bibinfo{person}{D. {Pompili}}.} \bibinfo{year}{2017}\natexlab{}.
\newblock \showarticletitle{Cooperative Hierarchical Caching in
  $5$\uppercase{G} Cloud Radio Access Networks}.
\newblock \bibinfo{journal}{\emph{IEEE Network}} \bibinfo{volume}{31},
  \bibinfo{number}{4}, \bibinfo{pages}{35--41}.
\newblock


\bibitem[\protect\citeauthoryear{{Weifeng}, {Mingqi}, {Jia}, {Siguang},
  {Lijun}, and {Jian}}{{Weifeng} et~al\mbox{.}}{2020}]%
        {Weifeng2020CachingSocialTrust}
\bibfield{author}{\bibinfo{person}{L. {Weifeng}}, \bibinfo{person}{Z.
  {Mingqi}}, \bibinfo{person}{X. {Jia}}, \bibinfo{person}{C. {Siguang}},
  \bibinfo{person}{Y. {Lijun}}, {and} \bibinfo{person}{X. {Jian}}.}
  \bibinfo{year}{2020}\natexlab{}.
\newblock \showarticletitle{Cooperative caching game based on social trust for
  \uppercase{D}$2$\uppercase{D} communication networks}.
\newblock \bibinfo{journal}{\emph{International Journal of Communication
  Systems}} \bibinfo{volume}{33}, \bibinfo{number}{9}, \bibinfo{pages}{e4380}.
\newblock


\bibitem[\protect\citeauthoryear{Yang and Leskovec}{Yang and Leskovec}{2012}]%
        {Yang2012Community}
\bibfield{author}{\bibinfo{person}{J. Yang} {and} \bibinfo{person}{J.
  Leskovec}.} \bibinfo{year}{2012}\natexlab{}.
\newblock \showarticletitle{Community-Affiliation Graph Model for Overlapping
  Network Community Detection}. In \bibinfo{booktitle}{\emph{IEEE International
  Conference on Data Mining}}. \bibinfo{pages}{1170–1175}.
\newblock


\bibitem[\protect\citeauthoryear{{Yang}, {Wu}, {Chen}, {Wang}, {Chen}, and
  {Yao}}{{Yang} et~al\mbox{.}}{2018}]%
        {yang2018locass}
\bibfield{author}{\bibinfo{person}{Y. {Yang}}, \bibinfo{person}{Y. {Wu}},
  \bibinfo{person}{N. {Chen}}, \bibinfo{person}{K. {Wang}}, \bibinfo{person}{S.
  {Chen}}, {and} \bibinfo{person}{S. {Yao}}.} \bibinfo{year}{2018}\natexlab{}.
\newblock \showarticletitle{{LOCASS}: Local Optimal Caching Algorithm with
  Social Selfishness for Mixed Cooperative and Selfish Devices}.
\newblock \bibinfo{journal}{\emph{IEEE Access}}  \bibinfo{volume}{6},
  \bibinfo{pages}{60--72}.
\newblock


\bibitem[\protect\citeauthoryear{{Zhang} and {Zhu}}{{Zhang} and {Zhu}}{2018}]%
        {Zhang2018Hierarchical}
\bibfield{author}{\bibinfo{person}{X. {Zhang}} {and} \bibinfo{person}{Q.
  {Zhu}}.} \bibinfo{year}{2018}\natexlab{}.
\newblock \showarticletitle{Hierarchical Caching for Statistical QoS Guaranteed
  Multimedia Transmissions over 5G Edge Computing Mobile Wireless Networks}.
\newblock \bibinfo{journal}{\emph{IEEE Wireless Communications}}
  \bibinfo{volume}{25}, \bibinfo{number}{3}, \bibinfo{pages}{12--20}.
\newblock


\bibitem[\protect\citeauthoryear{{Zhang}, {Li}, {Luan}, {Fu}, {Shi}, and
  {Zhu}}{{Zhang} et~al\mbox{.}}{2019}]%
        {Zhang2019MobilityAware}
\bibfield{author}{\bibinfo{person}{Y. {Zhang}}, \bibinfo{person}{C. {Li}},
  \bibinfo{person}{T. {Luan}}, \bibinfo{person}{Y. {Fu}}, \bibinfo{person}{W.
  {Shi}}, {and} \bibinfo{person}{L. {Zhu}}.} \bibinfo{year}{2019}\natexlab{}.
\newblock \showarticletitle{A Mobility-Aware Vehicular Caching Scheme in
  Content Centric Networks: Model and Optimization}.
\newblock \bibinfo{journal}{\emph{IEEE Transactions on Vehicular Technology}}
  \bibinfo{volume}{68}, \bibinfo{number}{4}, \bibinfo{pages}{3100--3112}.
\newblock


\bibitem[\protect\citeauthoryear{{Zhu}, {Zhi}, {Chen}, and {Zhang}}{{Zhu}
  et~al\mbox{.}}{2017}]%
        {Zhu2017SociallyMotivated}
\bibfield{author}{\bibinfo{person}{K. {Zhu}}, \bibinfo{person}{W. {Zhi}},
  \bibinfo{person}{X. {Chen}}, {and} \bibinfo{person}{L. {Zhang}}.}
  \bibinfo{year}{2017}\natexlab{}.
\newblock \showarticletitle{Socially Motivated Data Caching in Ultra-Dense
  Small Cell Networks}.
\newblock \bibinfo{journal}{\emph{IEEE Network}} \bibinfo{volume}{31},
  \bibinfo{number}{4}, \bibinfo{pages}{42--48}.
\newblock


\end{thebibliography}
\end{document}